%% file: main.tex
\documentclass[runningheads]{llncs}
\def\ANON{0}

\usepackage{tikz}
\usepackage{amsmath}
\usepackage{amssymb}
\usepackage{graphicx}
\usepackage{algorithm}
\usepackage{algorithmicx}
\usepackage{algpseudocode}
\usepackage{epstopdf}
\usepackage{rotating}
\usepackage[square,sort,comma,numbers]{natbib}
\usepackage{xurl}
\usepackage{color}
\usepackage{fancybox}
\usepackage{pstricks}
\usepackage{balance}
\usepackage{pst-plot}
\usepackage{array}
\usepackage{multirow}
\usepackage{mathtools}
\usepackage{pstricks-add}
\usepackage{flexisym}
\expandafter\def\expandafter\UrlBreaks\expandafter{\UrlBreaks\do\_}
\usepackage{listings}
\usepackage{tabularx,booktabs}
\usepackage{xcolor}

\definecolor{codegreen}{rgb}{0,0.6,0}
\definecolor{codegray}{rgb}{0.5,0.5,0.5}
\definecolor{codepurple}{rgb}{0.58,0,0.82}

\lstdefinestyle{mystyle}{
	commentstyle=\color{codegreen},
	keywordstyle=\color{magenta},
	stringstyle=\color{codepurple},
	basicstyle=\ttfamily,
	breakatwhitespace=false,         
	breaklines=true,                 
	captionpos=b,                    
	keepspaces=true,                 
	showspaces=false,                
	showstringspaces=false,
	showtabs=false,                  
	tabsize=2
}

\usepackage{url}
\usepackage[colorlinks=true,citecolor=purple]{hyperref}
\hypersetup{colorlinks=true}

\let\origtau\tau
\renewcommand{\tau}{\scalebox{1.44}{$\origtau$}}

\ifnum\ANON=0
\newcommand{\SV}[1]{\textcolor{red}{{\sf (SVNote:}{\sl{#1}}{\sf)}}}
\newcommand{\SM}[1]{\textcolor{blue}{#1}}
\else
\newcommand{\SV}[1]{}
\newcommand{\SM}[1]{}
\fi

\newcommand{\etal}{\textit{et al. }}

\newcommand\blfootnote[1]{%
  \begingroup
  \renewcommand\thefootnote{}\footnote{#1}%
  \addtocounter{footnote}{-1}%
  \endgroup
}

\begin{document}

\date{}

\title{Passive Triangulation Attack on ORide}

\ifnum\ANON=0

\author{
Shyam Murthy\inst{1}\orcidID{0000-0002-0222-322X}  \and
Srinivas Vivek\inst{2}\orcidID{0000-0002-8426-0859}}
\institute{shyam.sm@iiitb.ac.in \and srinivas.vivek@iiitb.ac.in}
\authorrunning{S. Murthy and S.Vivek}
\else
\author{}
\institute{}
\fi
\maketitle
\ifnum\ANON=0
\blfootnote{
This version of the contribution has been accepted for publication, after peer review (when applicable) but is not the Version of Record and does not reflect post-acceptance improvements, or any corrections. The Version of Record is available online at: http://dx.doi.org/10.1007/978-3-031-20974-1\_8. Use of this Accepted Version is subject to the publisher’s Accepted Manuscript terms of use https://www.springernature.com/gp/open-research/policies/accepted-manuscript-terms
\\S. Murthy and S. Vivek are with the International Institute of Information Technology Bangalore, India. Corresponding Author: Shyam Murthy, shyam.sm@iiitb.ac.in}
\fi
\input{abstract}
%
\input{introduction}
\input{background}
\input{ourwork}

\input{exptResults}

%
%
\input{relatedwork}
\input{conclusion}

\appendix
\section{Algorithms for triangulation attack}
\label{appendix}
\input{algo_appendix}

\bibliographystyle{splncs04}
\bibliography{abbrev0,crypto,morerefs}

\end{document}

%% file: abstract.tex
\begin{abstract}
Privacy preservation in Ride Hailing Services is intended to protect privacy of
drivers and riders.  ORide is one of the early RHS proposals published at USENIX
Security Symposium 2017.  In the ORide protocol, riders and drivers, operating in a zone,
encrypt their locations using a Somewhat Homomorphic Encryption scheme (SHE) and 
forward them to the Service Provider (SP).  SP homomorphically computes the 
squared Euclidean distance between riders and available drivers.  
Rider receives the encrypted distances and selects the
optimal rider after decryption.  In order to prevent a triangulation attack, SP randomly
permutes the distances before sending them to the rider.

In this work, we use propose a passive attack that uses triangulation to determine
coordinates of all participating drivers whose permuted distances are 
available from the points of view of multiple honest-but-curious adversary riders.
An attack on ORide was published at SAC 2021.  The same paper proposes a countermeasure using noisy Euclidean distances to thwart their attack.
We extend our attack to determine locations of drivers when given their permuted and
noisy Euclidean distances from multiple points of reference, where the noise perturbation
comes from a uniform distribution.

We conduct experiments with different number of drivers and for different perturbation
values.  Our experiments show that we can determine locations of all drivers 
participating in the ORide protocol.  For the perturbed distance version of the ORide protocol,
our algorithm reveals locations of about 25\% to 50\% of participating drivers.
Our algorithm runs in time polynomial in the number of drivers.

\begin{keywords}
	Ride Hailing Services, Privacy and Censorship, Attack, ORide, Triangulation.
\end{keywords}
\end{abstract}

%% file: introduction.tex
\section{Introduction}

Ride-Hailing Services (RHS) and Ride-Sharing Services (RSS) have become popular
world over.
RHS are typically managed by big organizations like Uber, Ola, etc.
RSS tend to be carpool or vanpool services offered by individuals in
a peer-to-peer car sharing model like Sidecar, Relay Rides, etc.
There are many advantages of both these services like being environment friendly
as well as being cost-effective.   In order to provision their services, RHS Service Providers (SP)
gather personal information from users, where users are both riders and drivers.  
Such information is needed for billing and statistics as well as to comply with regulatory requirements.
This might lead to a breach of privacy either due to negligence on part of the SP 
\cite{Zhao2019GeolocatingDA}, \cite{careem_rhs} or by the SP itself trying to snoop
user information for advertising or other purposes.
Therefore, privacy concerns of users
need to be addressed in order make RHS/RSS secure.  

A number of works are available in the literature that deal with privacy preservation
in RHS (PP-RHS).  While rider privacy concerns have received a lot of attention in PP-RHS literature, there are not many works that deal with driver privacy.  
There have been many instances in the media where driver safety is compromised.
The article \cite{thejournal} reports of how a gang  makes use of ride-hailing apps to harvest driver locations for robbery.  Another article \cite{istanbul_uber}, claims that some regular (non-SP) taxi drivers, pretending to be customers, located Uber vehicles to attack them.  These incidents provide a motivation to consider driver-threat models that leak locations of drivers to riders.

ORide \cite{ORidePaper} is one of the early RHS proposals which primarily aims to provide 
secure and efficient ride matching between riders and participating drivers.
ORide and many subsequent PP-RHS solutions make use of homomorphic encryption
schemes as their primary cryptographic construct.  
Homomorphic encryption schemes allow computation
on ciphertexts such that the result of computation is available only after decryption.
A Fully Homomorphic Encryption (FHE) scheme allows homomorphic addition and 
multiplication any number of times at the cost of large ciphertexts and huge 
computation latency.   A Somewhat Homomorphic Encryption (SHE) scheme
also allows addition and multiplication but only for a pre-defined number of times
with the advantage of much smaller ciphertext size and lesser computational latency.  
Hence, SHE schemes are more popular in practical solutions, e.g., ORide in which the FV SHE scheme
\cite{EPRINT:FanVer12} is used to provide 112-bit security.
Most of the PP-RHS solutions use semantically secure encryption schemes in their
protocol.  However, possible side-effects of adapting these encryption schemes to RHS applications
need to be considered carefully which otherwise might give scope for
adversarial attacks.  

In the ORide protocol, described in more detail in Section \ref{sec:oride}, rider and drivers rely on SHE to send their
encrypted locations to the SP.  SP then homomorphically computes the encrypted
squared Euclidean distances between rider and drivers.  These encrypted distances are received
by the rider who decrypts the distances, selects the smallest among them and informs
the index of the selected driver.  SP notifies the driver and helps establish a
secure channel between rider and the selected driver after which both the parties proceed
with ride completion.   The paper considers drivers' location-harvesting attacks where
a malicious rider might try to triangulate drivers by making three simultaneous fake
ride requests and obtain distances.  They propose a countermeasure where SP permutes 
the list of drivers' indices before forwarding to rider.  This is aimed to thwart the ability
of the adversary to correlate driver distances and hence perform triangulation.
However, we show that four colluding adversary riders, by using only the uncorrelated 
driver distances available to each of them as per the ORide protocol, and using
triangulation, can reveal the coordinates of all the participating drivers.

Alongside the RHS proposals, a number
of attacks on some of the works are also available in the literature, a few of them
are mentioned in Section \ref{sec:rel_works}.
We note here that
the attacks are important to help make the protocols more robust and thus aid in
development of truly privacy-preserving versions of RHS protocols.
The work by Kumaraswamy \etal \cite{deepakOridePaper} proposes a location-harvesting
attack on the ORide protocol, where they show that a rider can determine the exact
locations of 45\% of the responding drivers.  They make use of the fact that the
planar coordinates in the ORide protocol are integers and obtain all lattice points
inside a circle of radius $n_d$ centered at rider's actual location, where
$n_d$ is the distance computed by rider between herself and driver $d$.
They then reduce the number of potential locations,
to about 2 on average, using topology information of the region and
by making use of the fact that a driver must be at a motorable location in rider's zone.
They propose a countermeasure to thwart this attack where the driver, instead
of encrypting her actual location, encrypts a uniform random location within a 
perturbation distance $\rho$ from her original location to respond to rider's request.  
In effect, using anonymised driver location will result in the rider decrypting a 
{\em noisy distance} between herself and the driver but all the other steps in the
original ORide protocol remain unchanged, and we term this as the {\em noisy}
variant of ORide.  
They show that by using such anonymised locations, drivers are successful in 
preserving their anonymity against their attack.

In this paper, we show that a sufficient number of adversary riders colluding with one
another and using only the {\em uncorrelated noisy distances} available to them, can
reveal locations of about 25\% to 50\% of participating drivers.  Each of our successfully
recovered locations will be a distance at most $2\rho$ from a driver's original location.
Our solution neither makes use of topology information of the region nor relies on
encoding mechanism of driver coordinates.  While the attack of \cite{deepakOridePaper} 
on the ORide protocol reveals coordinates of about 45\% of participating drivers, 
our attack on the same reveals the coordinates of all participating drivers.

\subsection{Triangulation}
Triangulation is a well-known method in navigation and surveying where the position of
an object can be determined based on angles measured from multiple points of reference.
The term triangulation has been used in different contexts.  
In the field of image processing and computer vision, the term triangulation refers to the
process of finding the position of a point in space by using its position in two images
taken with different cameras \cite{hartley1997}.
In computational geometry, polygon triangulation is a basic algorithm where a polygon
is decomposed into a set of non-intersecting triangles \cite{siedel1991}.
In this paper, we use the term triangulation to mean determining the planar coordinates
of an object based on distances measured from multiple reference points.

\subsection{Threat Model}
As mentioned earlier, users in RHS can either be drivers who sign up with the SP to offer rides for profit
or riders who wish to avail ride services for a fee.
Either of the three entities, namely the SP, driver or rider can be an adversary
trying to learn more about other parties.
The SP, holding personal user information, has a reputation to maintain and hence
would typically refrain from taking an active adversarial role, except that it is curious.   
Riders can be subjected
to profiling or even physical attacks, where drivers are the adversary entities.
Riders or other outsiders masquerading as legitimate riders can be the adversary entities
trying to obtain driver information like location, identity etc.   In the ORide protocol,
riders and drivers are modeled as active adversaries and SP is an honest-but-curious 
adversary.  Riders and drivers do not collude with SP and that the SP does not provide the 
users with malicious apps.

In this work, we consider the adversarial model as that of ORide.
We look at driver location-harvesting attacks by adversary rider entities.
In other words, the rider is the adversary trying to glean location information about
participating drivers and the SP is unaware of presence of such an adversary.
The adversary, in this case,
will follow the protocol correctly but uses the information gathered during the protocol to
infer more than what is ordinarily available in the protocol.
Multiple adversary riders participate in the attack, share information with one another 
and thus collude among themselves to perform driver location-harvesting attack. 
In a practical scenario, a competitor SP can masquerade as a set of riders with the intention 
of gleaning location information about drivers working for other SP(s).

\subsection{Our Contribution}
We present two driver location-harvesting attacks, one on the ORide protocol
and another on the variant of the ORide protocol suggested as countermeasure
in the attack paper of \cite{deepakOridePaper}.

Our first attack algorithm, given in Section \ref{sec:attack_oride}, makes use of 
triangulation to harvest drivers' locations.
The ORide protocol claims to preserve driver anonymity in addition to providing
an efficient and privacy-preserving ride hailing protocol.  In particular, they
claim to provide a countermeasure for driver triangulation attack by malicious riders.
This is done by randomly permuting driver responses before forwarding them to
requesting riders, thus preventing the rider adversaries from correlating distances
between the adversary and responding drivers.
Our attack uses four honest-but-curious passive adversary riders in collusion,
where each adversary gets uncorrelated drivers' distances in random permuted order.
We experimentally show that our algorithm is successful in obtaining exact coordinates of all 
drivers who respond to ride requests, thus
negating one of the security claims of the ORide paper.
Our attack recovers coordinates of all responding drivers, by picking points that
are at exact distances received by atleast three adversaries, and is independent of
the underlying road topology, whereas the attack in \cite{deepakOridePaper}
uses road information and recovers about half the number of responding drivers.

We next consider an attack on the {\em noisy} variant
of the ORide protocol which was proposed in \cite{deepakOridePaper} as a countermeasure
mechanism for a topology-based driver location-harvesting attack proposed in the 
same paper.  In the countermeasure solution, each rider obtains uncorrelated and noisy 
(squared) distances 
to anonymised locations within a circle of radius $\rho$, the perturbation distance, around each of the responding drivers, which is the noisy variant of the ORide protocol.  
In Section \ref{sec:attack_noisy_oride}, we provide an algorithm to recover
locations of responding drivers in the noisy variant of the ORide protocol.
We use 12 colluding adversaries in this attack where each adversary is honest-but-curious and passive,
all of whom receive permuted $n$ noisy distances from $n$ participating drivers.
Using $n$ such distances each from two adversaries, our attack looks at all intersecting $2n^2$ points and 
then does a brute force over distances available with all other adversaries and eliminates 
unsuitable points.
We run our experiments for different values of
$\rho$ and for different zone sizes as described in Section \ref{sec:expRes}.   
Our results are tabulated in Tables \ref{tab:results1} and \ref{tab:results2} which show that
we obtain coordinates of about 25\% to 50\% of participating drivers.
Each of our successfully recovered coordinates will be at a distance at most $2\rho$
from a driver's original location.

Our algorithms are heuristic in nature and each runs in time $\mathcal{O}(dn^3)$, for
$d$ adversaries and $n$ drivers participating in the protocol.
Our attack can result in large-scale inference of
drivers' locations which might lead to unwanted consequences.
We note here that our attack does not reveal anything about the driver private
information other than their location.  Our attacks are independent of the
underlying encryption scheme and the coordinate encoding mechanism.

The rest of the paper is organized as follows. Section \ref{sec:oride} describes
the ORide protocol and its noisy variant.   Section \ref{sec:our_work} explains the attacks followed by our experiments and results in Section \ref{sec:expRes}.
Related works in Section \ref{sec:rel_works} gives some details of literature that identify vulnerabilities in 
PP-RHS solutions.
The algorithm for the two attacks are given in Appendix \ref{appendix}.

%% file: background.tex
\section{Overview of ORide \cite{ORidePaper} and its Noisy Variant \cite{deepakOridePaper}}
\label{sec:oride_noisy}
In this section, we begin with a brief overview of ORide and present only those details that are needed to demonstrate our attacks.  Next, we describe the {\em noisy} variant
of the ORide protocol which mitigates a location-harvesting attack from \cite{deepakOridePaper}.
\subsection{ORide \cite{ORidePaper}}
\label{sec:oride}
ORide  is a privacy-preserving ride-hailing service protocol.
The protocol is summarized in the following steps:
\begin{enumerate}
\item There are three parties involved in the protocol : rider, service provider (SP) and drivers.
\item  SP does not collude with rider or drivers.  Drivers and riders
are honest-but-curious adversaries trying to learn more about the other party than what
is expected from the protocol.  The protocol aims to preserve the identity and location information
of drivers and riders from SP and from each other.
\item  During initialization, SP creates and publishes zones in its area of operation.  
Zones are demarcated based on historical driver and ride densities ({\it aka} anonymity set) in the region and such information is available to all drivers and riders alike.
\item   Rider, wanting to hail a ride, picks a (public key, private key) pair for a semantically 
secure SHE scheme, encrypts
her locations using the public key and sends the ciphertext and public key to SP requesting a ride.    The authors use the FV SHE scheme \cite{EPRINT:FanVer12} to demonstrate the
working of the protocol.
Since the basic SHE scheme (without HEEAN/CKKS extensions \cite{CheonKKS17}) 
works on integers, locations are encoded as integers in UTM\footnote[1]{Universal 
Transverse Mercator: a map-projection system for geographical locations \cite{UTMGrid}} format. 
\item  SP forwards the request to all drivers in rider's zone.
\item  Available drivers encrypt their locations using the public key and send the ciphertext to SP.
\item  SP does not have the secret keys to decrypt the received ciphertexts.
The semantic security of the underlying SHE scheme ensures the privacy of the encrypted data with respect to SP.
\item  SP uses the homomorphic properties of the SHE scheme to securely compute the squares of 
the Euclidean distances between each of the received drivers' encrypted locations and 
the encrypted rider location.
\item  SP randomly permutes the results before forwarding them to rider.
This is done to prevent correlation of driver responses by rider and thus helps to prevent
triangulation attack by multiple colluding riders trying to derive driver locations.
\item  Rider uses the SHE private key to decrypt the distances, sorts the resulting plaintext,
chooses the index associated with the smallest value and informs the SP.
\item  SP enables the rider and selected driver to establish a secure communication channel 
between each other.  Following this, the rider and driver complete their ride establishment
privately.
\item  In addition to the aforementioned features, the protocol also provides convenience 
features like ride payment, reputation rating, accountability etc., which are
not relevant for the attack in our paper.  The interested reader is referred to
the original paper for more details.
\end{enumerate}

In Section 8 of the ORide paper, the authors consider location-harvesting attacks by outsiders.
They consider triangulation attack on drivers where the adversary can obtain driver
coordinates by making three simultaneous ride requests from different locations.  
They consider two methods to mitigate the attack.  
The first one is to make it financially expensive, wherein they impose a
charge on riders who issue ride requests followed by cancellations.  The second method
is to permute the list of drivers' indices for each ride request so as to thwart the
ability of the adversary to correlate rider request with driver responses.
In this paper, we provide an attack on the second mitigation method mentioned above.  

\subsection{Noisy Variant of ORide \cite{deepakOridePaper}}
\label{sec:noisy_ver}
The paper \cite{deepakOridePaper} proposes a location-harvesting attack on
the ORide protocol.   Their attack describes a method
where a rider in ORide, by issuing a single ride request, is able to determine the 
exact locations of about 40\% of the responding drivers.   Their method uses
the topology, terrain and road network information of the area to mount the attack.
Their method is a variant of the classical Gauss' circle problem of finding lattice
points on a circle of known radius.  They find integer coordinates on a 
circle of radius equal to the distance between rider and responding driver, with
center of circle being the rider location.  They then reduce the number of possible locations
by using road network information by picking points that lie only on motorable areas.
In the same paper, in order to thwart the attack mentioned therein, the authors
propose a simple countermeasure where each driver anonymises her location by choosing a random
location within a distance $\rho$ from her original location.  
In other words, in response to a ride request, a driver instead of encrypting her
actual location, would pick a random anonymised location inside a circle of radius $\rho$
around her actual location 
and sends the encryption of that anonymised location.   The value of $\rho$
is determined by the ride-hailing application provided by SP 
based on observed driver density in the specific zone, available to both drivers and riders.
SP then homomorphically computes the squared Euclidean distance
between the rider location and the anonymised location and forwards the same to rider.
After decryption, rider would thus obtain a noisy distance between herself and the 
responding driver.

The maximum distance between the actual location and the random anonym-ised location is the perturbation 
distance $\rho$.  Due to this perturbation, the authors of \cite{deepakOridePaper} show a marked increase in driver anonymity
with respect to rider.    Note that increasing the perturbation distance would decrease
accuracy of selected closest driver, thereby requiring the SP to select the perturbation
distance carefully to prevent customer dis-satisfaction.  Finally, the authors give 
recommendation for perturbation distance based on grid size.
We term this variation of ORide as {\em noisy} ORide.

We also give an attack on noisy ORide to try and recover coordinates of participating
drivers.

%% file: ourwork.tex
\section{Location Harvesting Attacks}
\label{sec:our_work}
In this section, we describe our location-harvesting attacks on the ORide protocol.
We start with a basic triangulation attack wherein the received responses can
be correlated to the respective drivers.
As mentioned in Section \ref{sec:oride}, the ORide protocol has a mechanism to thwart 
such an attack; so we propose an attack which is an adaptation of the basic triangulation
attack to recover driver coordinates of all
participating drivers in the ORide protocol.  We then look at the attack on the noisy version of
ORide.  Our attack, in this case, will try to recover a point within the perturbation circle
around the driver's actual location.
The algorithms of our attack are given in Appendix \ref{appendix} as Algorithms \ref{algo:AlgoMainZP} 
and \ref{algo:AlgoMain}.

\subsection{Preliminaries}
We consider the honest-but-curious adversary model in our attack.
The adversary rider follows the protocol and tries to obtain more
information based on driver response than is ordinarily available as per the protocol.
SP does not collude with any of the parties.  Multiple adversaries participate in the attack
and all of them collude among themselves.
RHS providers operate in specific geographic areas and for ease of 
management, the SP demarcates the area into suitably sized zones.
Based on their location, the SP associates riders and drivers to a specific 
zone for purposes of ride hailing and ride availability advertisements, respectively.

\label{subsec:basics}
\vspace{0.05in}
\noindent {\bf Definition: Adversary Circle:}  Adversary circle w.r.t adversary $A_\alpha$, 
at location $\alpha$, and $i\textsuperscript{th}$ driver $D_i$,
denoted by $\mathcal{K}_{\alpha, R_{\alpha, i}}$, is a circle centered at $\alpha$ with 
radius $R_{\alpha, i}$, the squared Euclidean distance between 
$A_\alpha$ and $D_i$ (at her original or anonymised location). \\
\noindent {\bf Definition: Perturbation Circle:} Perturbation circle w.r.t a driver's (actual or potential) location 
$\chi$, denoted by $\mathcal{P}_{\chi, n\rho}$, is a circle centered at $\chi$ 
and radius equal to $n\rho$ ($n \in \mathbb{R})$, and $\rho$ is the perturbation distance.
In the ORide protocol, driver would set $n$ to 0 since there is no perturbation being added
to her original location.
In the noisy variant of ORide, driver would set $n=1$ when responding to ride request.
Adversary rider would set $n=2$ while executing the procedures from Algorithm \ref{algo:AlgoMain}.

\noindent {\em Remarks:}
\begin{enumerate}
\item None of the attacks described in this paper make use
of any geographical information or road network topology in the area of operation.   
\item The location and identity of the entity requesting a ride are encrypted, hence the SP
will not be able to distinguish between an adversary ride request from a legitimate one.
\item Each adversary entity is indistinguishable from one another.  In a practical attack,
there might be just one single physical device masquerading as multiple adversaries.
\item  UTM coordinates are
used by the ORide protocol since they use integer based BFV-SHE scheme \cite{EPRINT:FanVer12} in their protocol.  ORide could possibly be 
adapted to use CKKS FHE scheme \cite{CheonKKS17} to encrypt
real number coordinates in (latitude, longitude) format.  
Our attack is agnostic of the underlying encryption scheme or coordinate scheme 
since it only uses Euclidean distances computed between drivers and adversaries. 
\end{enumerate}

\subsection{Basic Triangulation Attack}
\label{sec:bta}
In the basic triangulation attack, three colluding riders located sufficiently close 
to one another issue ride requests simultaneously.  
Each rider computes the distance values using responses obtained from the SP,
for each responding driver.  The drivers' locations can be computed easily
as long as the distances corresponding to a specific driver can be correlated 
with respect to each of the riders.  If $A_\alpha$, $A_\beta$ and $A_\gamma$
be three adversary riders, $R_{\alpha, i}$, $R_{\beta, i}$
and $R_{\gamma, i}$ be distances computed for driver $D_i$ by each of
the adversaries, respectively, then the
point of intersection of $\mathcal{K}_{\alpha, R_{\alpha, i}}$, 
$\mathcal{K}_{\beta, R_{\beta, i}}$ and $\mathcal{K}_{\gamma, R_{\gamma, i}}$
would give the location of driver $D_i$.
The ORide paper specifically
mentions a possible triangulation attack and provides a mechanism to thwart the 
same as described in \ref{sec:oride}.
In the following section, we describe our attack to overcome the same.

\subsection{Attack on ORide Protocol} 
\label{sec:attack_oride}
As described in Section \ref{sec:oride}, ORide provides a mechanism to thwart the basic rider triangulation attack.  They prevent the ability of adversary rider to correlate
the computed distances without which the basic triangulation attack described
in Section \ref{sec:bta} would not be possible.
Our attack uses triangulation and involves multiple adversaries placed far away from one another.
Our algorithm runs in time $\mathcal{O}(dn^3)$, for $d$ adversaries and $n$ drivers.
In our experiments, four adversary entities were enough to reveal all responding drivers' locations.  
Each adversary entity issues a ride request (almost) simultaneously following the ORide protocol.  
SP then computes distances homomorphically from the received responses, permutes the
order of received responses and forwards the set of encrypted distances to each requester,
respectively.  Hence, each adversary obtains $n$ distances in random permuted order.
Each adversary colludes with one another to execute Algorithm \ref{algo:AlgoMainZP}.
A brief description of the algorithm is given below.

\subsubsection{Description of Algorithm \ref{algo:AlgoMainZP}} 
\begin{itemize}
\item Let $A_\alpha$ denote the adversary rider, where $\alpha \in [1,4]$.  For $n$ 
responding drivers, each $A_\alpha$ would receive $n$ distances.
\item Two adversary entities, say, $A_1$ and $A_2$ w.l.o.g, each use their respective
$n$ distance values to form $\mathcal{K}_{1, R_{1, i}}$ and
$\mathcal{K}_{2, R_{2, i}}$, for $i \in [1,n]$, respectively.
\item $A_1$ and $A_2$ collude and intersect $\mathcal{K}_{1, R_{1, i}}$, for $i \in [1,n]$, and
$\mathcal{K}_{2, R_{2, j}}$, for $j \in [1,n]$, to obtain a set $L$ of $2 n^2$ intersection points.  
This is done in the procedure {\tt ObtainCircleIntersectionPoints}.
\item Each of the other two adversary entities $A_3$ and $A_4$ use their respective $n$
distance values to form $\mathcal{K}_{3, R_{3, i}}$ and
$\mathcal{K}_{4, R_{4, i}}$, for $i \in [1,n]$, respectively.
\item $A_3$ and $A_4$ each check if points in the $L$ set lie on their respective circles.
\item Points that do not lie on either of the circles are filtered out from $L$.  This is done in  procedure {\tt FilterInCorrectCoordinates}.  
\item At the end of this procedure only valid drivers' coordinates
 are present in the set $L$.
\end{itemize}

\subsection{Attack on Noisy Version of ORide}
\label{sec:attack_noisy_oride}
As described in Section \ref{sec:noisy_ver}, \cite{deepakOridePaper} gives a 
location-harvesting attack on ORide paper and a countermeasure solution for their attack.  
In this section, we give an attack on the solution proposed in that paper.
Our attack uses a variation of triangulation and involves multiple adversaries placed far away from one another.
Our algorithm runs in time $\mathcal{O}(dn^3)$, for $d$ adversaries and $n$ drivers.
In our experiments, we use 12 adversary entities for the attack.

Formally, let $A_i$ be the $i\textsuperscript{th}$ adversary rider, $i \in [1,12]$ at 
location $(x_{A_i}, y_{A_i})$.
Each $A_i$ issues a ride request (almost) simultaneously following the ORide protocol.   
As part of the attack mitigation given in \cite{deepakOridePaper}, a driver 
at location $(x_D, y_D)$, upon receiving the request, picks a uniformly random location
$(\hat{x}_{D_i}, \hat{y}_{D_i})$
inside a perturbation circle of radius $\rho$ centered at $(x_D, y_D)$, 
encrypts that random location and responds to the ride request.
SP receives the response, uses it to homomorphically
compute the squared Euclidean distance between $(x_{A_i}, y_{A_i})$ and $(\hat{x}_{D}, \hat{y}_{D})$ and forwards the same to $A_i$.  $A_i$ then uses the secret 
key to decrypt the computed distances to obtain the noisy
distances between itself and each of the responding drivers.
The adversaries collude with one another and execute Algorithm \ref{algo:AlgoMain}.
The steps of the algorithm are described below.

\textit{Remark :} The distances received by each adversary are in random order of drivers.  Hence, the subscript $j$ in $R_{\alpha, j}$ is just an enumeration and not a driver identifier.

\subsubsection{Description of Algorithm \ref{algo:AlgoMain}} 
\begin{itemize}
\item Each adversary $A_\alpha$, for $\alpha \in [1,m]$, is placed at a location far away from one another inside the zone, and each one receives noisy Euclidean distances from each of the participating drivers $D_i$, for $i \in [1,n]$. 
\item Let $R_\alpha = \{R_{\alpha, 1}, \ldots, R_{\alpha, n}\}$ be the list of distances
received by adversary $A_\alpha$ from $n$ drivers in some random order.
\item Two adversary entities, say, $A_1$ and $A_2$ w.l.o.g, each use their respective
$n$ distance values to form $\mathcal{K}_{1, R_{1, i}}$ and
$\mathcal{K}_{2, R_{2, i}}$, for $i \in [1,n]$, respectively.
\item $A_1$ and $A_2$ collude and intersect $\mathcal{K}_{1, R_{1, i}}$, for $i \in [1,n]$, and
$\mathcal{K}_{2, R_{2, j}}$, for $j \in [1,n]$, to obtain the set $L_1$ of $2 n^2$ intersection points, $n$ (to be determined) of which are potential points inside each driver's perturbation circle.
This is done in the procedure {\tt ObtainCircleIntersectionPoints}.
\item For driver locations that are close to being collinear with locations of $A_1$ and $A_2$,
there is a high probability that the adversary circles do not intersect.
In order to retrieve such points, we pick two other adversaries $A_3$ and $A_4$ located at points that are non-collinear to locations of both $A_1$ and $A_2$.
\item $A_3$ and $A_4$ collude and intersect $\mathcal{K}_{3, R_{3, i}}$, for $i \in [1,n]$, and
$\mathcal{K}_{4, R_{4, j}}$, for $j \in [1,n]$, to obtain the set $L_2$ of $2 n^2$ intersection points as mentioned above.
\item We pick the first point $p$ from $L_1$.  We draw perturbation circle $\mathcal{P}_{p, 2\rho}$.  (The radius of the perturbation circle
is set this way since two points
in the driver's perturbation circle of radius $\rho$ will be at a distance  at most 
$2 \rho$ from one another.) 
\item Using each of the $n$ distances available at next adversary $A_\alpha$, 
we look for intersection of $\mathcal{P}_{p, 2\rho}$ and
$\mathcal{K}_{\alpha, R_{\alpha, i}}$, for $i \in [1,n]$.
If none of the 
$\mathcal{K}$ circles intersect $\mathcal{P}_{p, 2\rho}$ at real points,
we discard $p$.  Otherwise, it is retained. This is done in procedure
\texttt{FilterOutSuperfluousPoints}.
\item The above step is repeated for each point in $L_1$ as well as in $L_2$ and over each
adversary circle of every adversary.
\item  At the end of the previous step, we are left with two sets of filtered points in $L_1$ and $L_2$.
\item The procedures \texttt{FilterOutNearbyInvalidPoints} and \texttt{SelectLikelyPoints}
take $\tau$ as an input parameter.  It is the threshold distance between two
points to filter out nearby points.
In case of small zone size and/or large number of drivers, we end up with a large number of points in $L_1$ and $L_2$ close to each other.  In such cases we set $\tau$ to $< 2\rho$ so as to
enable filtering out more number of points.
\item Within each $L_i$, $i \in [1,2]$, we discard points that are at a distance $\le \tau$
from one another.    In procedure \texttt{FilterOutNearbyInvalidPoints}, we find such
points and pick only one of such nearby points.  
\item Finally, in the procedure \texttt{SelectLikelyPoints}, we compute the distance between
each point in $L_1$ with each point in $L_2$ and pick only those points in $L_1$ that are 
at a distance $\le \tau$ from at least one point in $L_2$.  These picked points
are output by the algorithm.
\end{itemize}

\subsection{Correctness and Time complexity}
The correctness of Algorithm \ref{algo:AlgoMainZP} is based on the fact that given
exact distances between driver and adversary locations, all the adversary circles intersect 
at the driver location.

\begin{figure*}[htpb]
\begin{center}
\includegraphics[width=3in]{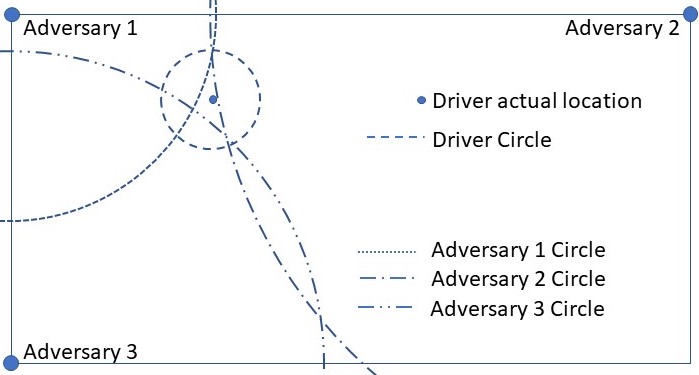}
\vspace{-.1in}
\caption{Grid showing three adversary circles intersecting within a driver circle. }
\label{fig:fig_adv}
\end{center}
\end{figure*}

Our attack in Algorithm \ref{algo:AlgoMain} relies on the fact that when given adversaries located
sufficiently far from one another and driver locations that are picked uniformly random
within the perturbation circle of radius $\rho$ centered around the respective driver
original location, 
then with a high probability the intersection of two such sets of adversary circles would 
have an intersection point inside the perturbation circle
of {\em some} driver, as shown in Figure \ref{fig:fig_adv}.  
Since the maximum distance between two points inside a circle of radius $\rho$ 
is $2\rho$, we use one such intersection point as center of a circle
of radius $2\rho$ and then look for the remaining adversary circles 
that will intersect this circle (of radius $2\rho$).

However, when two adversary locations and a driver's actual location are
close to being collinear then there is a high probability that the circles w.r.t that
particular driver do not intersect.
Hence,
we pick two other adversaries that are located in such a way so as to obtain successful 
intersection of the adversary circles corresponding to the {\em missed} drivers. 

Algorithms \ref{algo:AlgoMainZP} and \ref{algo:AlgoMain}, given in Appendix \ref{appendix} and described in 
Section \ref{sec:our_work}, each run in time $\mathcal{O}(dn^3)$, for $d$ adversaries and $n$ drivers.

\subsection{Mobile drivers in the ORide protocol}
\label{sec:mobile}
In Sections \ref{sec:bta} and \ref{sec:attack_oride}, we consider the case of a driver 
at a location $(x_D, y_D)$.
It is possible that a driver in the ORide protocol looking to offer a ride
might be mobile on the road.  In this section, we consider such non-static drivers who 
respond to ride requests.

We feel it is reasonable to consider the driving speed of a driver looking for a ride to
be about $25~kmph$.  This would translate to about 7 meters per second.
Assuming that the average turnaround time between ride request and response
is less than 10 seconds, the driver would have travelled about 70 meters.
This would imply that the actual location of the driver would be off by 
at most $\pm 70~m$ from the derived location obtained as described in Section \ref{sec:attack_oride} which is a reasonable approximation of driver's actual location.
This problem of harvesting locations of mobile ORide drivers can also be seen as a
problem in the noisy ORide variant where the perturbation $\rho$ is about $75m$.
Hence, we can apply Algorithm \ref{algo:AlgoMain} to obtain locations of mobile
drivers in the ORide protocol.

%% file: exptResults.tex
\section{Experiments and Results}
\label{sec:expRes}
Sagemath 8.6 \cite{sagemath} was used to implement the procedures mentioned in Algorithms \ref{algo:AlgoMainZP} and
\ref{algo:AlgoMain}.   Our experiments were run on a commodity HP Pavilion laptop with
512 GB SSD and AMD Ryzen 5 processor.   Our experiments take about 2 seconds
to complete for 25 drivers and about 300 seconds to complete for 100 drivers.
The Sagemath code for our attacks in Algorithms \ref{algo:AlgoMainZP} and \ref{algo:AlgoMain}  are 
available at \href{https://github.com/shyamsmurthy/cans2022}{https://github.com/shyamsmurthy/cans2022}.

\begin{table}[htpb]
\begin{center}
  \renewcommand{\arraystretch}{1.2}
  \begin{tabularx}{200pt}{c|c}
    \hline
Zone size &
Number of drivers = 25, 50, 75, 100 \\
 \hline
25 $km\textsuperscript{2}$ &  100  \\
100 $km\textsuperscript{2}$ & 100   \\
 \hline
\end{tabularx}
\caption{Percentage of driver locations recovered for the ORide protocol (zero perturbation)}
\label{tab:results0}
\end{center}

\begin{center}
  \renewcommand{\arraystretch}{1.2}
  \begin{tabularx}{275pt}{c|X|X|X|X|X}
    \hline
    \multirow{2}{*}{\centering Number of Drivers} &
      \multicolumn{5}{c}{\centering Perturbation radius (in meters)} \\
 &  \centering $\rho$ = 50 & $\rho$ = 75 & $\rho$ = 100 & $\rho$ = 125 &  $\rho$ = 150 \\
 \hline
25 &  60.8 & 59.7 & 51.6 & 49.3 &42.8  \\
50 & 50.1 & 43.8 &  42.5 & 35.4 & 33.3 \\
75 &  46.9  & 35.7 & 30.9 & 27.9  & 28.2\\
100 & 38.7  & 29.1 & 25.1 & 25.2 & 24.8   \\
 \hline
\end{tabularx}
\caption{Percentage of driver locations recovered for zone size of 100 $km\textsuperscript{2}$}
\label{tab:results1}
\end{center}

\begin{center}
 \renewcommand{\arraystretch}{1.2}
\begin{tabularx}{220pt}{c|X|X|X}
    \hline
    \multirow{2}{*}{\centering Number of Drivers} &
      \multicolumn{3}{c}{\centering Perturbation radius (in meters)} \\
 &  $\rho$ = 50 & $\rho$ = 75  & $\rho$ = 100 \\
 \hline
25 &  49.4 & 43.2  & 40.5 \\
50 & 38.5 &  35.4  & 30.0  \\
75 &  31.3  & 27.8 & 27.5 \\
100 & 24.9  &  24.5 & 24.4  \\
\hline
\end{tabularx}
\caption{Percentage of driver locations recovered for zone size of 25 $km\textsuperscript{2}$}
\label{tab:results2}
\end{center}

\begin{center}
\begin{tabular}{c | c | c}
\hline
Adversary count & Zone size = 100 $km\textsuperscript{2}$ & Zone size = 25 $km\textsuperscript{2}$ \\
\hline
	4 &  6.1  & 6.6\\
	8 &  22.8  & 18.7 \\
	\bf{12} &  \bf{46.9} & \bf{31.3}  \\
	16 & 50.5 & 26.0 \\
\hline
\end{tabular}
\caption{Percentage of driver locations recovered for different adversary count for perturbation distance = 50 meters and number of drivers = 75}
\label{tab:results4}
\end{center}
\end{table}

\subsection{Experiment Details}
Our experiments were run for two zone sizes, 100 $km\textsuperscript{2}$ and
25 $km\textsuperscript{2}$, which we term as large and small zone sizes, respectively.   
For the large zone size,
we ran experiments for perturbation values, $\rho$, of 0, 50, 75, 100, 125 and 150 meters.
Perturbation value of 0 corresponds to our attack described in Section \ref{sec:attack_oride}
and others correspond to our attack in Section \ref{sec:attack_noisy_oride}.
For the small zone, we have considered perturbation values, $\rho$, of 0, 50, 75 and 100 meters. 
These values are inline to what is mentioned in Section 3.2 of \cite{deepakOridePaper},
where the perturbation values are so chosen as to maintain sufficient anonymity and also to provide good accuracy
of driver selection in a given zone size.
For each combination of (zone, $\rho$) we set $n$, the number of drivers, to 25, 50, 75 and 100. 

At the start of each run of the experiment, the algorithm takes the zone coordinates and generates $n$
uniformly distributed random points, namely the driver original locations, inside the zone.  
For a driver original location $\chi$, and per adversary, a new uniform random point $p$ is 
picked at a distance $\le \rho$ from $\chi$. 
The squared Euclidean distance between $p$ and adversary location 
is given to the adversary in random order.  For $n$ drivers, each adversary will have $n$ distance values.
As is evident, this exactly mimics the rider scenario in the ORide protocol where, after decryption,
the rider receives the squared Euclidean distances to each driver from herself.
As described in Section \ref{sec:attack_oride}, each adversary will collude with one another
and execute the steps described in Algorithm \ref{algo:AlgoMain} to obtain a set
of $\eta$ coordinates.

\subsection{Discussion of results}
In our experiments for the attack on ORide protocol described in Section 
\ref{sec:attack_oride}, we exactly recover $n$ locations corresponding to each
of the $n$ drivers participating in the ORide RHS protocol and are given in Table \ref{tab:results0}.

The results from our experiments to recover driver locations
for the attack described in Section \ref{sec:attack_noisy_oride}, namely {\em noisy} ORide, are
given in Tables \ref{tab:results1} and \ref{tab:results2}.  Each table gives the
percentage of driver coordinates that are {\em valid}, averaged over 20 runs.
By valid, we mean that the recovered driver coordinates are at a distance $\le 2\rho$  from an original 
driver location.  Determining the set of valid locations is done during post-processing of
the results as explained in Section \ref{postproc}.   We see that as the number of
drivers increases, the number of valid locations revealed decreases.  This is due to
the fact that the number of circle intersections increases as square of number of drivers,
hence, resulting in a large number of false positives, thus decreasing the accuracy of the results.
In our experiments, we use $\tau$ as a parameter, to filter out 
values in procedures \texttt{FilterOutNearbyAlliedPoints}
and \texttt{SelectLikelyPoints} in Algorithm \ref{algo:AlgoMain}.

Table \ref{tab:results4} gives the percentage of successfully recovered driver locations,
out of 75 participating drivers and perturbation distance of 50 meters, for varying number
of adversary entities of 4, 8, 12 and 16.  We see that with 12 adversaries we get an optimum
recovery percentage.  With lesser number of adversaries, fewer points get eliminated resulting
in lower success rate.  With higher number of adversary circles and with a smaller grid size, 
far too many points get eliminated also resulting in a lower success rate.
Although, with zone size of 100 $km\textsuperscript{2}$ the recovery percentage 
is better for 16 adversaries, the gain is not proportional to the extra overhead, hence
we choose to keep the number of adversaries as 12.

\begin{table}[htpb]
\begin{center}
  \renewcommand{\arraystretch}{1.2}
  \begin{tabular}{c|c|c|c}
    \hline
	   & Success percentage of & Topology & Integer encoding \\
	   & attack on ORide Protocol & Information & \\
 \hline
	  Attack of \cite{deepakOridePaper} & 40 & Required & Required \\
	  Our Attack & 100 & Not Required & Not Required \\
 \hline
\end{tabular}
\caption{Comparison of our attack with the attack of \cite{deepakOridePaper}}
\label{tab:compare1}
\end{center}
\end{table}

\begin{table}[htpb]
\begin{center}
  \renewcommand{\arraystretch}{1.2}
	\begin{tabular}{p{1.05in}|l}
  \hline
   &  Attack Conditions  \\
  \hline
	ORide paper \cite{ORidePaper}\newline (Section 8) &  Rider can correlate driver with respective exact distance \\
  Our Attack\newline  (Algorithm \ref{algo:AlgoMainZP}) &  Drivers' exact distances are randomly  permuted by SP \\
  Our Attack \newline (Algorithm \ref{algo:AlgoMain}) &  Drivers' perturbed distances are randomly permuted by SP \\
 \hline
\end{tabular}
	\caption{Summary comparison of our attacks with triangulation attack considered in ORide paper \cite{ORidePaper}}
\label{tab:compare2}
\end{center}
\end{table}

In Tables \ref{tab:compare1}  and \ref{tab:compare2}, we give a summary comparison of
our attacks with that of \cite{deepakOridePaper} and the triangulation attack considered in the
ORide paper \cite{ORidePaper}, respectively.

\subsection{Post-processing of our experiment results}
\label{postproc}
At the end of each run, the experiment will return $\eta$ 
number of coordinates, where $\eta$ may or may not be equal to $n$.  
As mentioned earlier, at the start of every run, $n$ original driver locations are generated.
During post-processing, we compare each of the $\eta$ values with $n$ and pick only
those that is at a distance $\le 2\rho$ from one of the $n$ values.  These picked values are deemed
{\em valid} driver locations.  \\
\textit{Remark :} To obtain the percentage of valid locations we divide 
the number of valid driver locations by $max(n, \eta)$.


%% file: relatedwork.tex
\section{Related Works}
\label{sec:rel_works}
\subsection{Ride-Hailing Services}
Zhao \etal\cite{Zhao2019GeolocatingDA} perform a systematic study of
leakage of drivers' sensitive data and show that large scale data harvesting of such
information is possible.
They analyze apps provided to drivers by Uber, Lyft 
and other popularly deployed SPs and show that a malicious outsider SP can harvest
driver data using the apps.
ORide is one of the early PP-RHS
solutions that not only uses a semantically secure SHE scheme to encrypt user and
driver locations to enable an SP to anonymously match drivers with riders, but also
supports RHS features like easy payment, reputation rating, accountability etc. 
Kumaraswamy \etal\cite{deepakOridePaper} give a location-harvesting attack on ORide.  Using a single ride request they demonstrate that a rider can reveal locations of up to 40\% 
of participating drivers.  Their attack makes use of the geography and road network
of the area to identify motorable areas to determine likely driver locations.  
They provide a countermeasure to thwart the attack while preserving sufficient anonymity.    

Wang \etal propose {\em TRACE} \cite{wangTrace}, a privacy-preserving dynamic
spatial query solution using bilinear pairing to encrypt locations of drivers and riders. 
Kumaraswamy and Vivek \cite{deepakIndocrypt} disprove the privacy claims of the TRACE
protocol.   Their attack exploits shared randomness used across different messages
and recovers masked plaintext used in the initial part of the RHS protocol.  
{\em lpRide} by Yu \etal \cite{lpRideYu} uses modified Paillier cryptosystem \cite{nabeelPaillier} for encrypting RNE transformed locations of riders and drivers.
All homomorphic distance computations and optimum driver selection operations 
are performed on a single SP server.
\cite{lpRideAttackSV} propose an attack on the modified Paillier scheme used in lpRide, allowing the service provider to recover locations of all riders and drivers in the region.
Vivek \cite{svivek_attack} gives an attack on the lpRide protocol.  
Their attack does cryptanalysis of the modified version Paillier 
cryptosystem and show that any rider/driver will be able to learn the
location of any other rider participating in the RHS. 
\subsection{Triangulation}
Triangulation is a popular technique used in recovering 2 and 3-dimensional structure of
an object using images from multiple points of reference.   This problem is non-trivial
in the presence of noise \cite{hartley1997}.  There are a number of works in the field of computer vision
that work on recovering points in space when given image coordinates perturbed by noise,
and using camera matrices
\cite{hartley1997}, \cite{sturm_triangulation}, \cite{mohr_triangulation}.
In the field of computational geometry, the term polygon triangulation
means decomposition of a polygon into a set of triangles \cite{baker_triangulation}, \cite{david_triangulation}, \cite{michelle_triangulation}.

%% file: conclusion.tex
\section{Conclusion and Future Work}
In this paper, we presented driver location-harvesting attacks on the ORide paper based on
passive triangulation.  We recover
coordinates of all participating drivers in the ORide protocol.  We also look at an attack
on a {\em noisy} version of the ORide protocol where we recover about 25\% to 50\% of
participating drivers' locations.  

One possible method to decrease the effectiveness of our attack might be for a driver to vary, within some known bounds, the
radius of her perturbation circle for each response.
This would force us to use the largest possible perturbation radius
in our attack algorithm, possibly reducing its effectiveness.  However, this might also
have an effect on the driver selection algorithm.  Therefore, both the countermeasure solution
and our attack on the same need to be experimentally evaluated, 
which we leave for future work.

As part of a future work, our method of adapting triangulation
attack can be explored for other protocols that make use of uncorrelated distances
for privacy preservation.  Our method and analysis are based on heuristics, a
theoretical analysis would be very useful to correlate successful driver location recovery
with perturbation radius. 

\subsubsection*{Acknowledgements.}
We thank the anonymous reviewers for their invaluable comments and suggestions,
which helped us improve the manuscript.
This work was partly funded by 
the INSPIRE Faculty Award (DST, Govt. of India) and the
Infosys Foundation Career Development Chair Professorship grant  for Srinivas Vivek.

%% file: algo_appendix.tex
\begin{algorithm}[!]
\caption{Retrieve coordinates using Triangulation}
\label{algo:AlgoMainZP}
\begin{algorithmic}[1]
\Procedure{Main}{$R_1, \ldots, R_m$} 
        \State \textbf{Input} :  $R_1, \ldots, R_m$: the list of distances received by adversaries $A_1, \ldots, A_m$ in that order
	\State \textbf{Output} : Set of likely driver coordinates
	\Statex
	\State $L \gets $  {\tt ObtainCircleIntersectionPoints}( $R_1, R_2$ ) 
	\For{$i \gets 3, m$}\\
	   \Comment{Use other adversary distances to filter in correct intersection points}
	   \State $L \gets$ {\tt FilterInCorrectCoordinates}( $R_i$, $L$ ) 
	\EndFor	
	\State \textbf{return} $L$
\EndProcedure
\Statex
\Procedure{ObtainCircleIntersectionPoints}{$R_\alpha, R_\beta$}
	\State \textbf{Input} :  $R_\alpha, R_\beta$: the list of distances received by adversaries $A_\alpha$ and $A_\beta$, respectively
	\State \textbf{Output} : Set of circle intersection points
	\Statex
	\State $L \gets null$
	\For{$i \gets 1, len(R_\alpha)$} \Comment{$len(X)$ gives number of elements in list $X$}
		\For{$j \gets 1, len(R_\beta)$}
			\State $p_1, p_2 = Circle\_Intersection\_Points(\mathcal{K}_{\alpha, R_{\alpha, i}}, \mathcal{K}_{\beta, R_{\beta, j}})$ 
		          \If {$p_1 \ne null$} 
				\State Add $p_1$ and $p_2$ to $L$
			\EndIf
		\EndFor
	\EndFor
	\State \textbf{return} $L$	
\EndProcedure
\Statex
\Procedure{FilterInCorrectCoordinates}{$R_\gamma, L$}
        \State \textbf{Input} :  $R_\gamma$: the list of distances received by adversary $A_\gamma$, $L$: current set of intersection points
	\State \textbf{Output} : Filtered set of coordinate points
	\Statex
	\State $L_{out} = null$
	\For{$i \gets 1, len(L)$}
		\For{$j \gets 1, len(R_\gamma)$}
			\If {$L[i]$ lies on $\mathcal{K}_{\gamma, R_{\gamma, i}}$} \Comment{Adversary circle, center : $\gamma$, radius : $R_{\gamma, i}$}
				\State Add $L[i]$ to $L_{out}$
			\EndIf
		\EndFor
	\EndFor
	\State \textbf{return} $L_{out}$	
\EndProcedure

\end{algorithmic}
\end{algorithm}

\begin{algorithm}[!]
\caption{Retrieve coordinates from noisy distances}
\label{algo:AlgoMain}
\begin{algorithmic}[1]
\Procedure{Main}{$R_1, \ldots, R_m$}
	\State \textbf{Input} :  $R_1, \ldots, R_m$: the list of distances received by adversaries $A_1, \ldots, A_m$, respectively
	\State \textbf{Output} : Set of likely driver coordinates
	\Statex \\ \Comment{Intersect adversary circles $\mathcal{K}_{1, R_{1, i}}$ and $\mathcal{K}_{2, R_{2, i}}$, for $i \in [1,n]$}
	\State $L_1 \gets$ {\tt ObtainCircleIntersectionPoints}( $R_1, R_2$ ) 	\For{$i \gets 3, m$}
           	\Comment{ Use all the other adversary distance lists }
	  	\State $L_1 \gets$ {\tt FilterOutSuperfluousPoints}( $R_i$, $L_1$ ) 
	\EndFor	\\
	\Comment{Intersect adversary circles $\mathcal{K}_{3, R_{3, i}}$ and $\mathcal{K}_{4, R_{4, i}}$, for $i \in [1,n]$}
	\State $L_2 \gets$ {\tt ObtainCircleIntersectionPoints}( $R_3, R_4$ ) 
	\For{$i \gets 1, 2, 5, m$}
	   \Comment{ Use all the other adversary distance lists }
	   \State $L_2 \gets$ {\tt FilterOutSuperfluousPoints}( $R_i$, $L_2$ ) 
	\EndFor

	\State $L_1 \gets$ {\tt FilterOutNearbyInvalidPoints}( $L_1$, $2 \rho$ ) 
	\State $L_2 \gets$ {\tt FilterOutNearbyInvalidPoints}( $L_2$, $2 \rho$ ) 
	\State $L \gets$ {\tt SelectLikelyPoints}( $L_1$, $L_2$, $2 \rho$ ) 
	\State \textbf{return} $L$
\EndProcedure
\Statex
\Procedure{ObtainCircleIntersectionPoints}{$R_\alpha, R_\beta$}
        \State{\textbf{Input} :  $R_\alpha, R_\beta$: the list of distances received by adversaries $A_\alpha$ and $A_\beta$ respectively.}
	\State{\textbf{Output} : Set of circle intersection points}
	\Statex
	\State $L \gets null$ 
	\For{$i \gets 1 , len(R_\alpha)$}
		\For{$j \gets 1, len(R_\beta)$}
			\State $p_1, p_2 \gets Circle\_Intersection\_Points(\mathcal{K}_{\alpha, R_{\alpha, i}}, \mathcal{K}_{\beta, R_{\beta, j}})$
		          \If {$p_1 \ne null$} 
				\State Add $p_1$ and $p_2$ to $L$
			\EndIf
		\EndFor
	\EndFor
	\State \textbf{return} $L$	
\EndProcedure
\Statex
\Procedure{FilterOutSuperfluousPoints}{$R_\gamma, L$}
        \State{\textbf{Input} :  $R_\gamma$: the list of distances received by adversary $A_\gamma$, $L$: Current set of filtered circle intersection points}
	\State{\textbf{Output} : Filtered set of circle intersection points}
	\Statex
	\State $resultList \gets null$

	\For{$i \gets 1, len(L)$}
		\For{$j \gets 1, len(R_\gamma)$}
			\State $q_1, q_2 \gets Circle\_Intersection\_Points(\mathcal{P}_{L[i], (2 \rho)}, \mathcal{K}_{\gamma, R_{\gamma, j}}  )$ \\	\Comment{Perturbation circle $\mathcal{P}_{L[i], (2 \rho)}$: Center : $L[i]$, Radius : $2 \rho$}

		          \If {$q_1 \ne null$}
				\State Add $L[i]$ to $returnList$; 
				\State \textbf{break}
			\EndIf
		\EndFor
	\EndFor
	\State \textbf{return} $resultList$	
\EndProcedure
\Statex
\algstore{part1}
\end{algorithmic}
\end{algorithm}

\begin{algorithm}
\begin{algorithmic}[1]
\algrestore{part1}
\Procedure{FilterOutNearbyInvalidPoints}{$L, \tau $}
        \State{\textbf{Input} :  $L$: the list of coordinate points, $\tau$: threshold distance to filter out points}
	\State{\textbf{Output} : Filtered set of coordinate points such that neighbouring points that are $\le$ threshold distance from points in the filtered set are removed  }
	\Statex
	\State $resultList \gets null$ 
	\For{$i \gets 1, len(L)-1$}
		\For{$j \gets i+1, len(L)$}
			\If { Euclidean\_distance$(L[i], L[j]) \leq \tau $ } 
				\State Add $L[i]$ to $resultList$, if not already present
			\EndIf
		\EndFor
	\EndFor
	\State \textbf{return} $resultList$
\EndProcedure
\Statex
\Procedure{SelectLikelyPoints}{$L_1, L_2, \tau$}
        \State{\textbf{Input} :  $L_1, L_2$: the list of filtered coordinate points, $\tau$: distance threshold for selection}
	\State{\textbf{Output} : Set of likely driver coordinate points}
	\Statex
	\State $L \gets null$ 
	\For{$i \gets 1, len(L_1)$}
		\For{$j \gets 1, len(L_2)$}
			\If { $L_1[i] == L_2[j]$ } 
				\State \textbf{continue} 
			\EndIf
			\If { Euclidean\_distance$(L_1[i], L_2[j]) \leq \tau$ } 
				\State Add $L_1[i]$ to $L$, if not already present
			\EndIf
		\EndFor
	\EndFor
	\State \textbf{return}  $L$	
\EndProcedure
\end{algorithmic}
\end{algorithm}